\documentclass{aa}

\usepackage{graphics,epsfig}

\begin{document}
\outer\def\gtae {$\buildrel {\lower3pt\hbox{$>$}} \over 
{\lower2pt\hbox{$\sim$}} $}
\outer\def\ltae {$\buildrel {\lower3pt\hbox{$<$}} \over 
{\lower2pt\hbox{$\sim$}} $}
\newcommand{\xmm} {{\sl XMM-Newton }}
\newcommand{\angs} {$\rm A$}
\newcommand{\+} {$\pm$}
\newcommand{\ergscm} {ergs s$^{-1}$ cm$^{-2}$}
\newcommand{\ergss} {ergs s$^{-1}$}
\newcommand{\ergsd} {ergs s$^{-1}$ $d^{2}_{100}$}
\newcommand{\pcmsq} {cm$^{-2}$}
\newcommand{\ros} {{\sl ROSAT }}
\newcommand{\exo} {\sl EXOSAT}
\def\rchi{{${\chi}_{\nu}^{2}$}}
\newcommand{\Msun} {$M_{\odot}$}
\newcommand{\Mwd} {$M_{wd}$}
\def\Mdot{\hbox{$\dot M$}}
\newcommand{\dg} {$^{\circ}$}

%
   \title{High Resolution Soft X-ray Spectroscopy of 
   M87 with the Reflection Grating Spectrometers on XMM-Newton}

   \author{
   I.~Sakelliou\inst{1} 
\thanks{
	Present Address:  
 	School of Physics \& Astronomy, University of Birmingham, 
	Edgbaston, Birmingham B15 2TT, UK} 
   J.R.~Peterson\inst{2}, 
   T.~Tamura\inst{3}, 
   F.B.S.~Paerels\inst{2},
   J.S.~Kaastra\inst{3}, 
   E.~Belsole\inst{4}, 
   H.~B\"{o}hringer\inst{5}, 
   G.~Branduardi-Raymont\inst{1}, 
   C.~Ferrigno\inst{3}, 
   J.W.~den Herder\inst{3},  
   J.~Kennea\inst{6}, 
   R.F.~Mushotzky\inst{7},
   W.T.~Vestrand\inst{8},
   D.M.~Worrall\inst{9} }

\offprints{I.Sakelliou}

\institute{$^{1}$ Mullard Space Science Laboratory,
University College London,
Holmbury St. Mary, Dorking, \\ Surrey, RH5 6NT, UK \\
$^{2}$ Columbia Astrophysics Laboratory, 550 W. 120th St., 
New York, NY10027, USA\\
$^{3}$ SRON National Institute for Space Research, Sorbonnelaan 2, 
3584 CA Utrecht, The Netherlands \\
$^{4}$ CEA Saclay, Service d'Astrophysique, 91191 Gif-sur-Yvette, France \\
$^{5}$ Max Planck Institut f\"{u}r Extraterrestrische Physik, \\
$^{6}$ Department of Physics, University of California, 
Santa Barbara, CA 93106, USA \\
$^{7}$ NASA/GSFC, Code 662, Greenbelt MD20771, USA\\
$^{8}$ NIS-2, MS D436, Los Alamos National Laboratory, USA \\
$^{9}$ Department of Physics, University of Bristol, 
Tyndall Avenue, Bristol BS8 1TL, UK \\}

\authorrunning{Sakelliou et al.}
\titlerunning{High Resolution Spectroscopy of M87}

\date{}

\abstract{ We present high-resolution X-ray spectroscopic observations
of M87 with the Reflection Grating Spectrometers on {\it XMM-Newton}.
We detect strong K-shell line emission from N, O, Ne, Mg, some
emission from He-like Si, a fully resolved set of Fe L-shell emission
spectra, and some emission from C.  The angular intensity
distributions of the strong emission lines are detectably resolved on
scales $(15-160) \arcsec$.  The gas in the inner arcmin of M87 has a
multi-phase structure, as indicated by the similarity of the emission
line profiles of Fe L shell ions with widely separated ionization
potentials. The global Fe L spectrum is approximately consistent with an
isothermal plasma at $kT_e \sim 1.8$ keV, in addition to a component
with a temperature distribution appropriate to an isobaric cooling
flow, but with a minimum temperature cutoff of $kT_{\rm min} \approx
600$ eV.  The behaviour of this cooling-flow component is
qualitatively similar to what is seen in other cooling flow
clusters. Finally, we do not find any strong evidence for a
spatial variation in abundances due to resonance scattering
redistribution in the inner arcminute of the core.

 }

\maketitle

\section{Introduction}

The giant elliptical galaxy M87, its active nucleus, and its halo have
been the subject of intensive studies at all wavelengths (see e.g.,
R\"oser \& Meisenheimer 1999). The extended diffuse emission has been
mapped in neutral atomic and molecular gas, and in the highly-ionised
hot gas that is revealed through X-ray emission.  X-ray observations
with the {\it Einstein} Observatory first found evidence for a mass of
rapidly cooling X-ray emitting gas at the core of the
system. Estimates for the density and temperature of this gas
suggested the presence of a `cooling flow' with a total mass
deposition rate of $\sim 10 M_{\odot}$ yr$^{-1}$ (Fabricant et
al. 1980; Stewart et al.  1984; Fabian et al. 1984). Direct
spectroscopic evidence for the presence of radiatively cooling gas was
obtained with the Focal Plane Crystal Spectrometer on {\it Einstein}
(Canizares et al. 1979) and the Solid State Spectrometer (Lea, et al. 
1982).

The advent of high-sensitivity X-ray imaging and spectroscopy with the
{\it Chandra} and {\it XMM-Newton} Observatories now provides the
opportunity for a detailed study of the physical conditions in the
cooling gas and its interaction with its environment. High spatial
resolution images of M87 have been obtained with {\it Chandra} (Wilson
\& Yang 2001).  Medium resolution {\it XMM-Newton} spatially resolved
spectroscopy has been published by B\"ohringer et al. (2001), Belsole
et al. (2001), Molendi \& Pizzolato (2001) and Matsushita et at
(2001).  For objects of moderate angular extent (up to approximately
1~arcmin), the Reflection Grating Spectrometers (RGS) on {\it
XMM-Newton} have the unique capability to provide high resolution
X-ray spectroscopy coupled with some spatial resolution in the
cross-dispersion direction. This combination of capabilities is
uniquely suited to conducting a sensitive study of the thermodynamic
properties of the cooling gas. Here, we describe preliminary results
of such a study, based on data obtained during the {\it XMM-Newton}
Performance Verification phase.

\section{Data Reduction and Analysis}

M87 was observed on June 19, 2000, for a total of $\approx 60$ ksec.
All instruments on board {\it XMM-Newton} were operating.  For a
description of the observatory and its instrumentation, we refer to
Jansen et al. (2001). The RGS has been described by den Herder et
al. (2001).

The data reduction was performed with the SAS (XMM-Newton Science
Analysis System), 
using the most recent calibration for
wavelength scale and effective area. Data reduction was performed with
{\sc rgsproc-0.91}; spectral response files were generated with 
{\sc rgsrmfgen-0.34}. Periods of high background were removed from the
data, leaving a net effective exposure time of $\approx 40$ ksec. The
total count rate in the first spectral order is about 1.3 counts
s$^{-1}$ over the $5-35$ \AA\ band (both spectrometers combined).

Spectra were extracted from both RGS images by selecting events that
lie within a rectangular mask with a width in the cross dispersion
direction of 30 pixels, or $1.1$ arcmin.  This angular width
encompasses the visible bulk of the emission detected in the RGS.  The
background in the RGS images within this aperture is almost negligible
(a few percent of the source signal). 
At the distance of M87 (17 Mpc; Freedman et al. 1994), the 1.1 arcmin
width corresponds to $\sim 5.4$ kpc, or a radius of 2.7 kpc. The
spatial resolution of the RGS in the cross-dispersion direction is
approximately 15 arcsec, so that our extraction region covers about
four spatial resolution elements. Along the dispersion direction, the
equivalent spatial resolution is about 20 arcsec. The bright core has
a diameter of order 10 arcsec, and we expect all spectral lines to be
slightly broadened: the broadening is $\sim 0.124 \times \theta$
(1$^{\rm st}$ spectral order; \xmm UHB), where $\theta$ is the source
extent in arcmin. For comparison, the wavelength resolution and
spectral resolving power of the spectrometers are $\sim$0.06 \AA, and $\sim$
250 at 15 \AA (FWHM) respectively. 

Of course, we also detect photons from all gas in those parts of the
halo that are projected onto our aperture. Due to the much larger
extent of the outer halo, it appears as a near-isotropic component in
the image, which means that halo photons appear as a pseudo-continuum
in our spectra. The intensity of this pseudo-continuum can be
suppressed by using the intrinsic energy resolution of the RGS focal
plane CCD cameras, and filtering the photons by matching each photon's
dispersion coordinate to its expected possible range of CCD pulse
heights. The sensitivity of the spectrometer to remaining 'out of
band' photons is automatically included when we calculate the response
of the RGS, based on the response to monochromatic light from an
extended source of given angular distribution.  We used the standard
cut on CCD pulse heights, which includes 90\% of the pulse height
response to monochromatic radiation (about $\sim 200$ eV wide).

Fig.~1 shows the first- and second- order spectra. As can be seen,
we detect strong Ly$\alpha$ emission from the hydrogen-like ions of
Mg, Ne, O, N, and C, in addition to emission from essentially
all species in the Fe L series (Fe XVII, or Ne-like, through Fe XXIV,
or Li-like). No significant He-like oxygen is detected (O VII
resonance, inter-combination, and forbidden lines at 21.60, 21.80, and
22.10 \AA).

\begin{figure*}
\begin{center}
\setlength{\unitlength}{1cm}
\begin{picture}(15,8)
\put(-2,14){\includegraphics{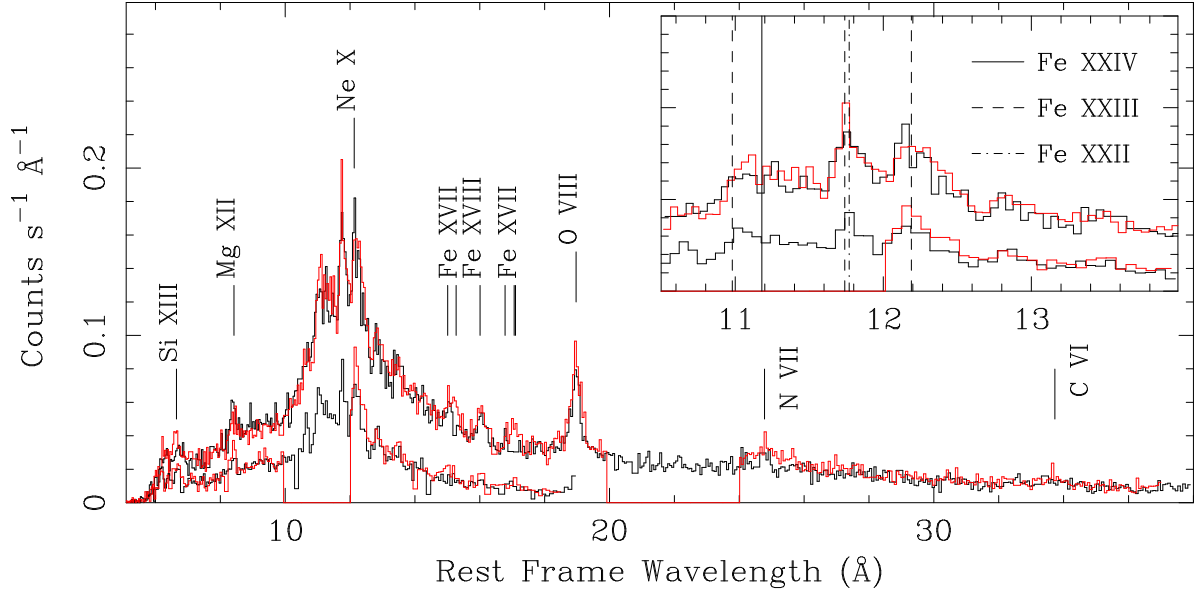}}
\end{picture}
\end{center}
\caption{
The total RGS spectrum of the central 1.1 arcmin of M87, in the source rest
frame. The spectra
from the two separate RGS instruments have been superimposed (RGS1 in
black, RGS2 in red). We also show the corresponding second order 
spectra (whose full range covers the $2.5-17.5$ \AA\ band). 
First and second order spectra have similar wavelength resolution, due
to the fact that the source is spatially resolved. Background has not
been removed. The positions of Ly$\alpha$ lines has been
marked, as well as those of the strongest lines of the two
lowest-ionization members of the Fe L series. The inset shows the
$10.5-14$ \AA\ region on an expanded scale, with the positions of the
strongest lines from the three highest-ionization members of the Fe L
series.
}
\end{figure*}

\section{Thermal Structure}

Spectroscopy in the $5-35$ \AA\ RGS range is particularly well suited
to studying the mix in temperature and abundance of cool gas in the
centre of M87. The presence of all eight charge states of the Fe L
series, with ionization potentials between 500 eV (to create the
Ne-like ion) and 2000 eV will allow us to use iron as a thermometer,
to probe the electron-temperature distribution between 200 and 2000
eV. There are no uncertainties associated with abundances, because the
temperature distribution is constrained solely by the ionization
balance of iron. Emission from the hydrogen-like ions of Mg, Ne, O,
and, to a lesser extent, N and C, can arise from gas at temperatures
outside the Fe L range, and therefore some uncertainties may remain in
the absolute abundances of these elements from the analysis of the RGS
spectrum.  Finally, because the core of M87 is moderately resolved in
the RGS on scales of order 15$\arcsec$, the analysis should also
provide constraints on the spatial temperature and abundance structure
below $kT_e \sim 2$ keV. However, this coupled spatial-thermal
structure of the core is complicated, and we therefore
chose to restrict the discussion in this Paper to a preliminary
analysis of the data in terms of a simple spatial model, based on the
directly observed image, convolved with a multi-temperature emission
spectrum.  A full characterisation of the RGS spectrum of an
extended source would require a more detailed theoretical model for the
spatial distribution of the extended source, and better knowledge of
the instrument and sky backgrounds.

Technically, the spectral analysis proceeded as follows. A
collisional-ionization equilibrium spectrum of chosen temperature
distribution and chosen abundance pattern was convolved with a model
that describes the spatial distribution of the emitting source.  This
spatial model was derived by using the \xmm MOS image in the relevant pulse
height range [(0.5-2.5)\,keV], properly scaled and collapsed onto the
RGS dispersion direction.  This spatial model was convolved with the
model emission spectrum, and compared to the data. Only the wavelength
range between 8 and 35\AA\, was used, since at the time of writing the
calibration accuracy below 8\AA\, was poorer, and therefore a 8\AA\, low
limit was a safe approach.  The flux normalisation of an extended
source in this procedure is such that it reproduces the counts
detected within the spatial extraction region. Therefore the numbers
we quote have not been corrected for the photons outside the
extraction region. Emission from the active nucleus, taken to be a
point source, was modelled with a power-law spectrum whose parameters
are based on our analysis of the EPIC spectrum (photon index $\Gamma =
2.3$, normalisation $7.9 \times 10^{49}$ photons s$^{-1}$ keV$^{-1}$
at 1 keV). Foreground absorption by our own Galaxy was modelled using
a column density $N_{\rm H,\ Gal} = 2.5 \times 10^{20}$ cm$^{-2}$. We
also allowed for intrinsic absorption at M87, modelled with a neutral
column density $N_{\rm H,\ intrinsic}$ that was optimised along with
the other free parameters, and fixed covering factor 0.5. A small
amount of photoelectric absorption by O intrinsic to the spectrometers
was modelled as absorption by a pure oxygen column density of $N_{\rm
O} = 3 \times 10^{17}$ cm$^{-2}$, based on observations of bright,
unabsorbed extragalactic continuum sources.

A background spectrum was extracted from RGS data of blank sky
(revolution 70). The background RGS
spectrum was extracted in a similar way as the source spectrum, and
scaled properly to match the RGS observation of M87. The contribution
from the background is relatively low as shown in Fig. 2.

Calculation of the model emission spectra and convolution with the
spatial distribution model were carried out using the {\sc spex}
package (Kaastra, Mewe, \& Nieuwenhuijzen 1996).  The assumption that
we can model the spatial distribution by convolving the emission
spectrum due to an on-axis point source with a suitably scaled angular
distribution (instead of properly taking the wavelength dependence of
the spatial response into account) is, in view of the compactness of
the source, only a minor one, and does not limit or affect our
conclusions.

The simultaneous appearance of widely separated (in ionization
potential) Fe L charge states already indicates that the core is not
isothermal; moreover, the shape of the emission features from
different ions suggests similar spatial profiles for different
temperature components. We therefore first attempted a model that
naturally allows for a multi-temperature and multi-phase structure,
the isobaric cooling-flow (ICF) prescription (Johnstone et al. 1992). Gas at
sub-keV temperatures, and at densities of \gtae $10^{-3}$ cm$^{-3}$,
has to radiatively cool rapidly, and would not be able to avoid
cooling to low temperatures if radiative cooling was the only term in
the thermal balance. However, spectroscopic observations of
essentially all cooling flows indicate that there is a very
significant lack of gas at low temperatures (Kaastra et al. 2001;
Peterson et al. 2001; Tamura et al. 2001). Adapting the procedure
followed by the above mentioned work on cooling-flow spectroscopy, we
introduce a minimum temperature to the isobaric cooling-flow
temperature distribution, in order to avoid 
overproduction of line emission in the lower few Fe L ions in a
cooling flow that correctly reproduces the emission from the higher
charge states.

\begin{figure*}
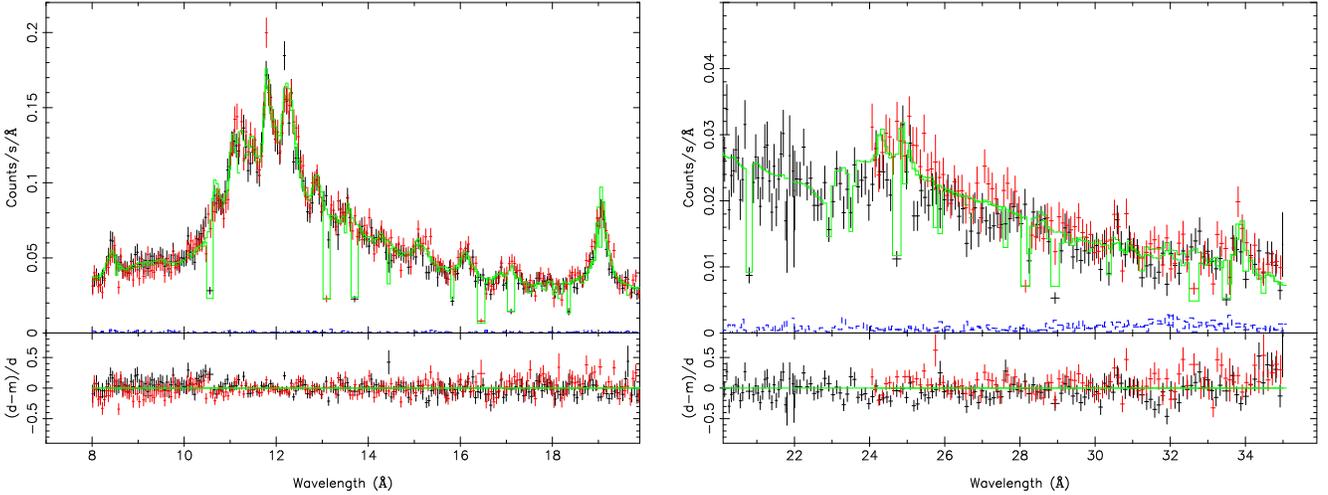

\begin{center}
\setlength{\unitlength}{1cm}
\begin{picture}(15,7)
\put(-2,7.5){\includegraphics{sakellioufig2a.ps}}
\put(7,7.5) {\includegraphics{sakellioufig2b.ps}}
\end{picture}
\end{center}
\caption{
The spectrum of the core of M87 fit with the 1T+ICF model, in 
the $8-20$
\AA\ range (left), and the $20-36$ \AA\ range (right). The RGS1
(black) and RGS2 (red) first order spectra were fitted simultaneously.
Green histogram is the best fitting model spectrum. The background
spectra employed in the modelling are shown in blue. The bottom panels
show the post-fit residuals. The large `dips' seen in both the data and the models correspond to CCD boundaries and/or dead columns. 
}
\end{figure*}

In agreement with the CCD spectral analysis (Finoguenov et al. 2001),
to this cooling flow we add a single temperature component (1T) to
account for emission from the hotter regions surrounding the core,
merging into the hot, extended intracluster medium. The
abundances of C, N, O, Ne, Mg, and Fe were left free, but tied to be
the same in both components. The abundances of all other elements were
set to their solar values relative to Fe; we do not expect that their
actual value effects the final results since the M87 RGS spectrum
[(8-35)\AA] is not sensitive to emission from elements such as Ca,
Ar, Al, Si, and S. Solar abundances were taken from Anders \& Grevesse
(1989).  The best-fit parameters are given in Table 1.

\begin{table}
\caption[]{Fitting the 1T+ICF model to the core of M87. 
Errors are statistical (90\% confidence level).}
\label{}
\begin{center}
\begin{tabular}{lcc}
\hline 
parameter			        & value \\
\hline 
$N_{\rm H,\ intrinsic}^a$		& $6\pm 2$ \\
covering factor$^a$ 			& $0.5 $\\
$\dot{M}^b$				& $6 \pm 1$\\	
$T_{\rm min}$ (keV)			& $0.6\pm 0.02$ \\
$EM_{\rm hot}^c$			& $8.1 $\\
$T_{\rm hot}$ (keV)			& $1.8\pm 0.05$\\
C$^d$					& $1.0\pm 0.3$\\
N$^d$					& $0.8 \pm 0.3$\\
O$^d$					& $0.49 \pm 0.04$\\
Ne$^d$					& $0.7\pm 0.2$\\
Mg$^d$					& $0.9\pm 0.2$\\
Fe$^d$					& $0.77\pm 0.04$\\

$\chi ^2/\nu$				& 962/694\\
\hline
\end{tabular}
\end{center}
\begin{description}
\item[$^a$] 'Intrinsic' column density in 
units of $10^{21}$~cm$^{-2}$, and the covering factor.
\item[$^b$] Mass deposition rate in 
units of ~\hbox{M$_{\odot}$}yr$^{-1}$.
\item[$^c$] Volume emission measure in units of $10^{64}$cm$^{-3}$.
\item[$^d$] Metal abundances relative to the solar values.
\end{description}
\end{table}

This '1T+ICF' model reproduces the observed spectrum reasonably well,
as is evident from Fig.~2.  An inspection of the relative
contributions of the isothermal component and the cooling flow
component to the spectrum (ICF : 1T $\sim$ 1 : 4) reveals that it is
actually dominated by the isothermal component; the main contribution
of the cooling flow are the faint Fe L lines between 12 and 16 \AA.
As stated above, emission from H-like ions, especially those of Mg,
Ne, and O, can also be produced by even hotter gas, which is not included
in our model. Evidence for hotter gas at larger radii comes from the
analysis of the observations from imaging detectors (e.g., Matsushita
et al. 2001).  Therefore, the derived abundances of those elements
have to be regarded with some skepticism.  The estimate for the total
mass deposition rate, $\Mdot$, is strongly coupled to that for the
minimum cooling flow temperature, for the simple reason that a high
$\Mdot$ over-predicts emission in the lowest Fe L ions, which has to
be countered by a relatively high minimum temperature. It is worth
noting that the previously morphologically determined deposition rate,
$\Mdot = 39^{+2}_{-9} M_{\odot}$ yr$^{-1}$ (Peres et al. 1998), when
scaled to our 1.1 arcmin aperture, predicts an $\Mdot \sim 7
M_{\odot}$ yr$^{-1}$.  Still, emission from cooling gas is not seen at
temperatures below $kT_e \sim 600$ eV, and all difficulties with
understanding this result in other cooling flow clusters apply equally
to the present case of M87. How strongly emission from cooling gas at
the lower temperatures is ruled out is illustrated in Fig.~3, where
we show the measured spectrum, overlaid with the '1T+ICF' spectral
model, but this time with the contribution of the cooling flow below
$kT_e = 600$ eV included.

\section{The Spatial
Distributions of Temperature and Abundance on Larger Scales}

Finally, we attempt to characterise the physical state of the line
emitting gas on spatial scales comparable to, or somewhat larger than
the spatial resolution of the RGS, by examining the intensity
distribution in a number of diagnostic, strong emission lines. Figure
4 shows their cross-dispersion profiles, binned in bins of 15 arcsec
in extent in the cross-dispersion direction, and accumulated over 0.2
\AA\ wide intervals in the spectral dimension . The contribution of
the continuum to the profiles has been subtracted, by constructing
cross-dispersion profiles in continuum windows, and interpolating the
strength of the continuum along the spectral dimension across the
emission lines.

The angular response of the RGS in the cross-dispersion direction was
calibrated on the spectral image of the bright BL Lac Mkn 421, which
is a point source (Brinkmann et al. 2001). Fig.~4(a) shows the angular
response to a point source in the cross-dispersion direction, overlaid
on cross-dispersion emission line profiles of M87, which shows that 
we do indeed resolve the core on \gtae $15\arcsec$ scales.

\begin{figure}
\begin{center}
\setlength{\unitlength}{1cm}
\begin{picture}(15,7)
\put(-0.5,7.5){\includegraphics{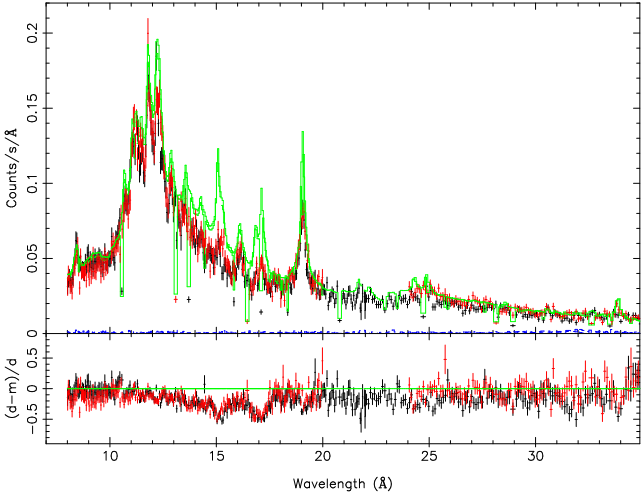}}
\end{picture}
\end{center}
\caption{The spectrum of the core of M87: the 1T+ICF model is shown
in green, but with all emission from the cooling flow originating from
gas below $kT_e = 600$ eV included.  The model over-predicts the
long-wavelength range of the spectrum. It strikingly over-produces line
emission from low-ionization Fe L ions, most notably Fe XVII ($2p-3d\
\lambda 15.014$ \AA, $2p-3s\ \lambda\lambda 17.051, 17.096$ \AA), and
from O VIII (Ly $\alpha\ \lambda 18.97$ \AA).  }
\end{figure}

The first two sets of profiles [in fig. 4(a) and 4(b)] were designed to test
for the possibility that the apparent spatial intensity distributions
of some of the strong emission lines are affected by radiative
transfer effects. Given the typical densities and scale sizes in the
cores of clusters, one estimates that the optical depth in
abundant-element emission lines with oscillator strengths of order
unity could be appreciable (of order unity), and resonance scattering
of these line photons would alter their angular intensity distribution
(Gilfanov et al. 1987).  Understanding this effect is
obviously important because it would complicate the interpretation of
these intensity distributions in terms of physical gradients in
temperature and abundance. In addition, constraints on individual
ionic column densities, when combined with volume emission measures in
the same ion, in principle over-constrain the emissivity model and
allow for important consistency tests. A byproduct of such
self-consistent over-constrained models is an absolute
(distance-independent) measurement of the important physical
parameters of the system. The first unambiguous detection (in the
sense that it is based on single-ion spectroscopy) of this effect was
recently reported by Xu et al. (2001) in the spectrum of the
elliptical galaxy NGC 4636.

Fig.~4(a) shows the profile of O VIII Ly$\alpha$, compared to the
corresponding Ly$\beta$ (the latter blended with Fe XVIII $\lambda
16.004$ \AA). The data indicate that the optical depth in O VIII
Ly$\alpha$ is low: there is no significant difference between the O
VIII Ly$\alpha$ and Ly$\beta$ profiles. It is expected that if
Ly$\alpha$ were optically thick, Ly$\beta$ would be optically thin,
due to its lower oscillator strength.  Note that there is a weak trend
for Ly$\beta$ to appear more diffuse than Ly$\alpha$, which is the
contradicts our expectetions if O VIII Ly$\alpha$ were optically
thick.  The same figure also shows the profile for a high-oscillator
strength transition in the Fe XVIII $\lambda 14.208$ \AA\ doublet. Its
profile is very similar to the O VIII Ly profiles.

Fig.~4(b) shows a similar set of profiles for high- and 
low-oscillator strength transitions in a low ionization Fe L ion, 
Fe XVII. The $\lambda 15.014$ \AA\ transition has the highest oscillator
strength of any line in the RGS band, yet its distribution looks similar
to that of the low-oscillator strength 
$\lambda\lambda 17.051, 17.096$ \AA\ blend, 
indicating that the optical depth is small. We conclude that the
core is optically thin in all important transitions.

Fig.~4(c) displays the profile observed in the $\lambda\lambda
11.176, 11.266$ \AA\ blend of Fe XXIV (unresolved due to the spatial
extent of the Fe XXIV emitting source), overlaid on the distribution
in Fe XVII $\lambda\lambda 17.051, 17.096$ \AA.  The fact that the
spatial distribution of the intensity in these two ions appears
qualitatively similar indicates that the `multiphase' character of the
gas we observe in the very core of M87, persists on scales out to
$\sim 150 \arcsec$, or \gtae $10$ kpc.

The last comparison of profiles is between the Ly$\alpha$ lines of
H-like O and Mg ($\lambda 18.97, 8.42$ \AA), 
and the resonance line of He-like Si ($\lambda 6.65$ \AA). The fact that 
these distributions appear similar indicates that there are no large
gradients in the relative abundances of these elements with respect to
each other. The O VIII Ly$\alpha$ profile appears slightly more peaked
than the others, which could simply be due to the fact that the
average gas temperature declines towards the very core.

\begin{figure*}
\begin{center}
\setlength{\unitlength}{1cm}
\begin{picture}(32,7)
\put(0.0,0.0){\includegraphics{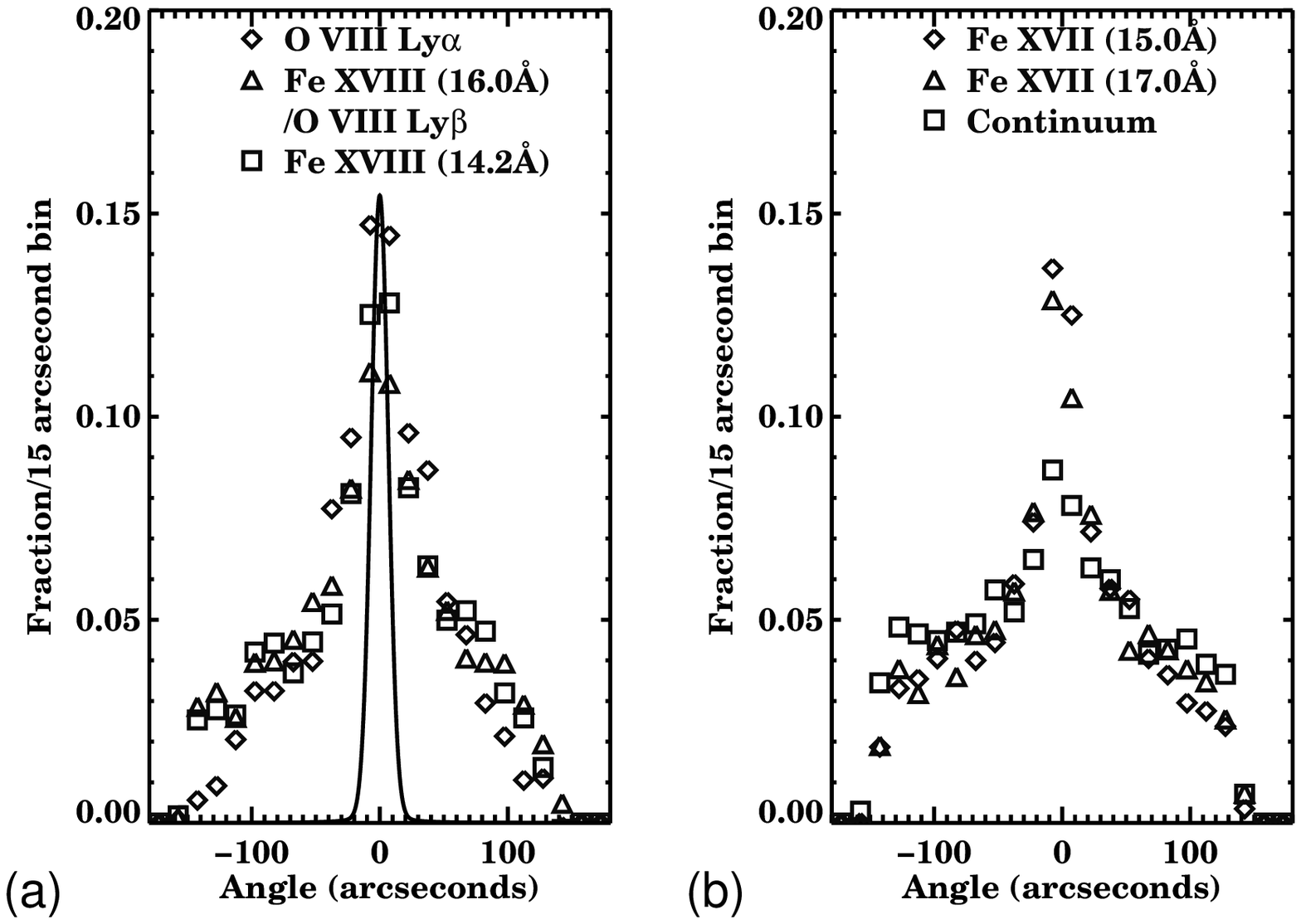}}
\put( 9.0,0.0){\includegraphics{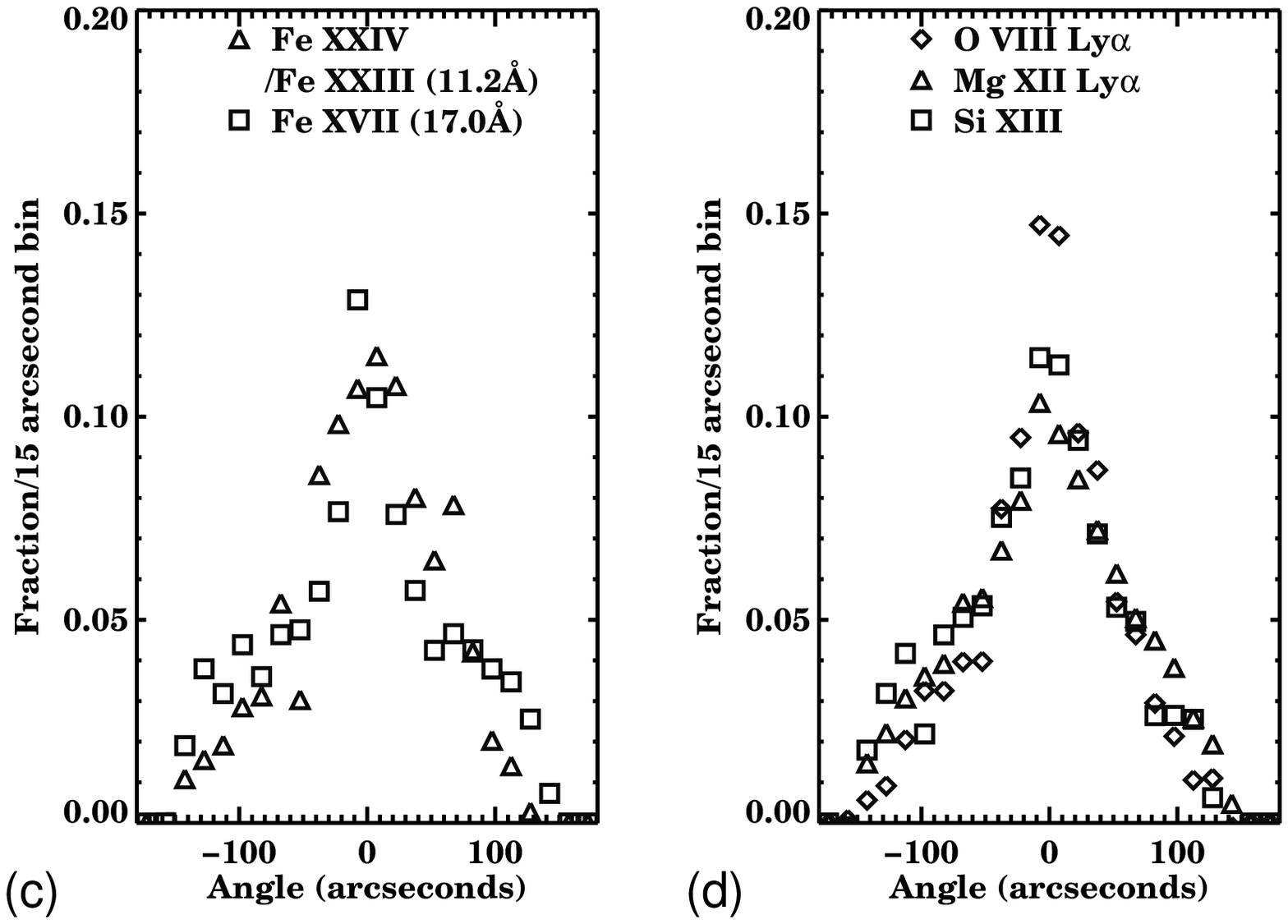}}
\end{picture}
\end{center}
\caption{Spatial intensity distribution in the cross dispersion
direction of a number of diagnostic emission lines, in 15 arcsec
bins. They were accumulated over $\pm$0.1\AA\, around the wavelength of
each line. The errors are smaller than the size of the symbols for the
core of the distribution, and double the size of the symbols for the
outer regions. The angular response of the RGS in the cross-dispersion
direction is shown in (a) (see \S4 for more information). }
\end{figure*}

\section{Summary and Discussion}

We have presented high resolution soft X-ray spectroscopy of M87. In
brief, the \xmm RGS spectrum shows fully spectrally resolved emission
from C, N, O, Ne, Mg, and Fe. The angular structure of the intensity
in the strong emission lines is resolved on scales of order \gtae 15
$\arcsec$, removing much of the confusion associated with the limited
spectral resolution of CCD spectrometers. We do not find any evidence
for resonance scattering in the central one arcminute. 

The gas in the inner $\sim 10$ kpc has a multiphase structure,
as indicated by the similarity of the spatial profiles of the emission
lines of Fe L ions with widely different ionization potentials. An
analysis of the overall ionization balance indicates that gas at all
temperatures down to $kT_{\rm min} \approx 0.6$ keV is present, with
an emission measure distribution approximately equal to that predicted for
isobaric radiative cooling.  Additionally, a strong approximately
isothermal component of $kT_e \approx 1.8$ keV is present, which gives
rise to emission from higher Fe ionization states (e.g., Fe XXIV), whose 
distribution is much wider than that of the
multiphased component, as seen in the CCD images (Molendi \& Pizzolato
2001; Matsushita et al. 2001).  The global deposition rate of the 
rapidly cooling gas is about $\Mdot \approx 6 M_{\odot}$ \, yr$^{-1}$ in
the inner arcmin.  In these respects, the core of M87 resembles that
of other cooling flow clusters. 

We do not find evidence for resonance scattering redistribution in the
strong emission lines. Detailed calculations of the effect of
radiative transfer in M87 (Mathews et al 2001) predict
that the line of sight average optical depth of the O VIII line is
$<$0.5, consistent with our results. However, the Fe XVII lines with
higher oscillator strengths are predicted to be marginally optically
thick in the absence of turbulence.  We do note that both our analysis
and the reanalysis of the EPIC data have reduced the apparent O
abundance deficit in the centre of M87, originally noted by
B\"{o}hringer et al. (2001); this reduces the need for an inclusion of
radiative transfer effects in the modelling of the M87 spectrum. If
anything, the accurate determination of the metal abundance depends on
the level of the continuum, which in turn depends on the contribution
from the active nuclear/jet emission. More confident measurements will
have to wait for data from which the non-thermal emission will be
subtracted, instead of modelled.

Resonance absorption scattering has been recently found in a poor
cluster of galaxies (NGC 4636; Xu et al. 2001). Apart from the apparent
differences between the two sources (e.g., density and temperature),
both of them host a radio galaxy. The radio galaxy in NGC~4636
(Birkinshaw \& Davis 1985; Stanger \& Warwick 1986) is much weaker
than that in M87: NGC~4636 has a steep spectrum, with a core power at
2~cm of $6.6 \times 10^{19} {\rm W \, Hz^{-1}}$ (Nagar et al.
2000). Both observational facts may imply that it is an old or a
`dead' radio source. However, as is becoming clearer, thanks to recent
numerical simulations (e.g., Br\"uggen et al. 2002), even `dead' radio
galaxies can stir up, disrupt the intergalactic medium, and introduce
turbulence. Therefore, a direct comparison between the two sources and
the high resolution X-ray results, is not straightforward and more
radio and X-ray observations with high angular resolution are required
to address these issues. It is expected though that turbulence in the
gas will reduce the net optical depth. Such turbulence could arise as
the radio jets impinge onto the intergalactic medium. We defer the
detailed discussion of the relative optical depths in the different
lines and the effects of turbulence to a later paper which will model such 
effects directly.

Our analysis of the RGS data of M87 finds near-solar abundances of C
and N, Ne and Mg, sub-solar for O, and 0.8 solar for Fe. We emphasise
that the absolute values for the abundances have to be regarded with
some caution, because they are based on fitting a parameterised model
with limited flexibility.  As pointed out by many authors (e.g.,
Matsushita et al. 2000, Buote 2000) CCD spectra are not of
sufficiently high resolution to obtain robust chemical abundances for
complex thermal plasmas. In particular if there are sharp abundance
gradients and multi-temperature gas the CCD data are degenerate, and
several solutions can exist with different temperature and abundance
values. The XMM CCD spectra for M87 (Finoguenov et al. 2002;
Gastaldello \& Molendi 2002) show both abundance gradients and
complex thermal structure and therefore their interpretation is open
to modelling uncertainties.  Both studies use rather different models
for the temperature distribution, but arrive at similar values for the
abundances except for oxygen. Gastaldello \& Molendi (2002) point out
that since the bulk of the emission in M87 is locally isothermal and
shows a rather smooth temperature gradient, the abundances derived
from the \xmm CCD data should not be very sensitive to the assumed
fitted temperature distribution; given the high signal to noise of the
XMM CCD data, all acceptable solutions should produce essentially the
same distribution of emission measure with temperature.

Our thermal modelling assumptions differ from those of both
Gastaldello \& Molendi (2002) and Finoguenov et al. (2002) but we do
not believe that they should result in different abundances. The
oxygen and Fe abundances we derive are the same as Finoguenov et
al. (2002) and Gastaldello \& Molendi (2002) averaged over the central
1~arcmin region.  However the abundances of Mg and Ne, which depend on
detailed modelling the Fe L shell lines in the CCD data (but not in
the RGS data) are rather different with the RGS results, both being
~(50-70)~percent higher. This comparison graphically illustrates the
difficulty of accurately measuring the abundances of these elements in
the temperature range for which the Fe L shell lines are strong.

The derivation of the Mg abundance is of particular importance: the
M87 stellar spectra show a very high apparent Mg/Fe ratio of \gtae 2:1
(Terlevich \& Forbes 2002) and a super-solar Mg abundance which is
completely incompatible with the RGS data of Mg/Fe $\sim$1 and Fe less
than solar.  Thus, very little of the observed X-ray emitting gas can
originate from stellar mass loss, contrary to the expectations of the
observed stellar density and normal stellar evolutionary theory. As
originally pointed out by Lowenstein \& Mathews (1991) a normal type I
supernova rate combined with stellar mass loss would produce a 3-5
times solar Fe abundance in the gas. As discussed in detail by Awaki
et al. (1994) the ASCA data did not detect such a high Fe abundance,
raising the serious issue as to what has happened to the products of
type I SN in the central regions of giant elliptical galaxies. These
results enhance and deepen the mystery. The sub-solar O, Ne, Mg and
Fe abundances are incompatible with the stellar data even if all the
heavy elements in the gas originate from stellar mass loss.

We also know that M87 is not unique. The abundance pattern in NGC~4636
 (Xu et al. 2001) is extremely similar and preliminary analysis of the
 RGS data for another giant elliptical NGC~533 (Peterson et al. 2002)
 shows similar results. We thus conclude that this issue is a serious
 one, and it needs further attention.
 
The \xmm RGS results pose two problems regarding the M87 chemical
abundances: i) why is the overall Fe abundance so low, and ii) why is
the chemical composition apparently so different from the stars.  We
speculate that the first problem can only be solved if either the true
type I SN rate is considerably reduced, or the yield of Fe per unit
supernova is also reduced. Otherwise, Fe may be lost from the central
regions of M87 [by ram pressure stripping for example, (Stevens et al.
1999)], or the ISM gas is diluted due to the accretion of cluster gas,
but both seem unlikely. The second issue can only be resolved if
either the inferred stellar abundances are in error, or if the stellar
spectra do not represent the true abundance distribution of the
stars. If for example, there are two distinctive stellar populations
in elliptical galaxies with different metal abundances, and the
optical data register only the abundances of the metal rich
group. Support for the chemical inhomogeneity of the stars in
elliptical galaxies comes from deep VLT and HST observations, of
NGC~5128 for example, (Harris \& Harris 2000) which shows a wide range
in metallicity of the stars.

\begin{acknowledgements}
This work is based on observations obtained with {\it XMM-Newton}, an
ESA science mission with instruments and contributions directly funded
by ESA Member States and the USA (NASA).
\end{acknowledgements}

\end{document}